\documentclass[prb,twocolumn,superscriptaddress,letterpaper]{revtex4}

\usepackage{amsfonts}
\usepackage{amsmath}
\usepackage{graphicx}
\usepackage{color}
\usepackage[colorlinks=true]{hyper ref}

\newcommand{\R}{\ensuremath{\mathbb R}}	
\newcommand{\GQ}{Q_\text{\rmfamily\itshape g}}
\newcommand{\GQbase}{q_\text{\rmfamily\itshape g}}

\renewcommand{\figurename}{{Figure}}
\renewcommand{\thefigure}{{\arabic{figure}}}
\renewcommand{\tablename}{{Table}}
\renewcommand{\thetable}{{\Roman{table}}}

\begin{document}

\title{Discovering Network Structure Beyond Communities}

\author{Takashi Nishikawa} 
\email[Correspondence and requests for materials should be addressed to T.N. (tnishika@clarkson.edu).]{}
\affiliation{Department of Mathematics, Clarkson University, Potsdam, NY 13699, USA}
\author{Adilson E. Motter}
\affiliation{Department of Physics and Astronomy and Northwestern Institute on Complex Systems, 
Northwestern University, Evanston, IL 60208, USA}
\affiliation{Department of Molecular Biology, Princeton University, Princeton, NJ 08544, USA}

\begin{abstract}
To understand the formation, evolution, and function of complex systems, it is crucial to understand the internal organization of their interaction networks.  Partly due to the impossibility of visualizing large complex networks, resolving network structure remains a challenging problem.  Here we overcome this difficulty by combining the visual pattern recognition ability of humans with the high processing speed of computers to develop an exploratory method for discovering groups of nodes characterized by common network properties, including but not limited to communities of densely connected nodes.  Without any prior information about the nature of the groups, the method simultaneously identifies the number of groups, the group assignment, and the properties that define these groups.  The results of applying our method to real networks suggest the possibility that most group structures lurk undiscovered in the fast-growing inventory of social, biological, and technological networks of scientific interest. 
\end{abstract} 

\onecolumngrid
\noindent
{\small Published in \href{http://www.nature.com/srep/2011/111109/srep00151/full/srep00151.html}{{\it Scientific Reports} {\bf 1}, 151 (2011)}; DOI:10.1038/srep00151}\\[5mm]

\maketitle

\begin{figure}
\includegraphics[width=\columnwidth]{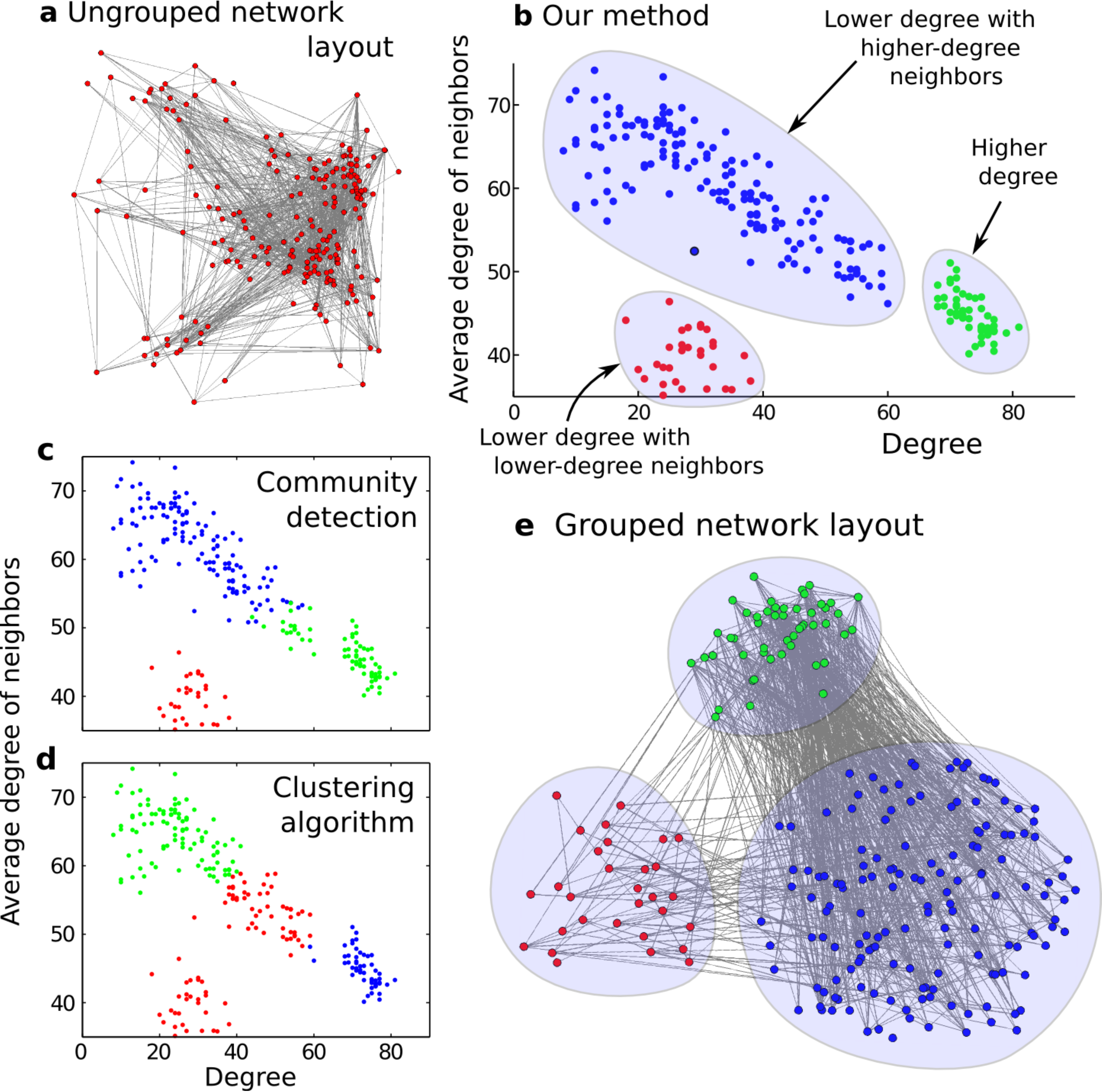}
\caption{\label{fig:example}
{\bf Discovering hidden group structure beyond density-based communities.}
\textbf{a}, Visualization of a network by the G\"ursoy-Atun algorithm~\cite{Gursoy2000dgslg}, which
attempts to place nodes uniformly while keeping the network neighbors close.
This and other standard layout algorithms fail to disentangle the network and reveal any clear group structure.
\textbf{b}, Using our visual analytics method, a user can discover three structural groups (of sizes 150, 50, and 30) without a priori information about the number of groups. 
The groups can be characterized by the degree and the neighbors' average degree, and 
at least two
properties are necessary to resolve the entire group structure.
\textbf{c}, Even the most general community detection method~\cite{Newman:2007rc} does not divide the network correctly. 
\textbf{d}, The $K$-means algorithm~\cite{Lloyd.:1982gf}, 
one of the most frequently used methods for general clustering problems, 
does not correctly capture the group structure when applied directly to the full node property space, even if the number of groups $K=3$ is given.
\textbf{e}, Layout of the network using the discovered groups.
For clarity, both panels \textbf{a} and \textbf{e} show only 10\% of the links.}
\end{figure}

The highly structured internal organization of complex networks can both impact and reflect their dynamics and function~\cite{Strogatz:2001il}.
Previous work on identifying and studying this organization has focused mainly on 
network communities~\cite{Girvan:2002fk,Radicchi:2004ve,Guimera:2004gf,Palla:2005fk,PhysRevE.74.016110,Danon:2006zr,Fortunato:2007fr,Chauhan:2009uq,Chen2010278,Mucha:2010fk,Porter:2009ht,Fortunato:2010uq}, which are subsets of nodes defined by the difference between their internal and external link density. 
To provide a fresh perspective on this problem, we seek to capture more general structures characterized by other network properties~\cite{Newman:2007rc,ravasz2002hierarchical,PhysRevE.75.036105,ISI:000286468600016}.  For this purpose, we introduce the notion of \textit{structural groups}, defined as subsets of nodes sharing common structural properties that set them apart from other nodes in the network. Using a given set of $p\!\gg\!\! 1$ node properties (such as centrality and spectral properties) as the coordinates for each node in the $p$-dimensional space $\R^p$, 
we identify structural groups as clusters of points in this \textit{node property space}. 
Figure~\ref{fig:example} shows an illustrative example of a network for which no standard network visualization shows clear group structure (Fig.~\ref{fig:example}\textbf{a}). However, an appropriate two-dimensional projection in the node property space reveals a hidden, but unambiguous three-group structure (Fig.~\ref{fig:example}\textbf{b}), which can be used to generate a far more informative
layout of the network (Fig.~\ref{fig:example}\textbf{e}). Application of existing community detection 
methods~\cite{Clauset:2005ly,Bagrow:2005mz,Raghavan:2007rt,Rosvall:2007ys,PhysRevLett.100.258701,Kovacs:2010qy,Wen:2011kx,Lancichinetti:2011yq,Estrada:2011fj,Psorakis:2011vn}
is not expected to resolve these groups, since they are not distinguishable by link density alone (Fig.~\ref{fig:example}\textbf{c}). 
Neither is the direct application of existing clustering methods in the full node property space nor in the projection onto any lower-dimensional space, due to the known fact that groups with widely different scatter sizes may not be correctly grouped by unsupervised algorithms (Fig.~\ref{fig:example}\textbf{d} and Supplementary Fig.~S1).  
Distinguishing structural groups may in general require a combination of  two or more properties --- Fig.~\ref{fig:example}\textbf{b} shows that the degree and the average degree of neighbors suffice for this example.  It is difficult, however, to identify such a combination without knowing the groups a priori.

\begin{figure*}
\includegraphics[width=\textwidth]{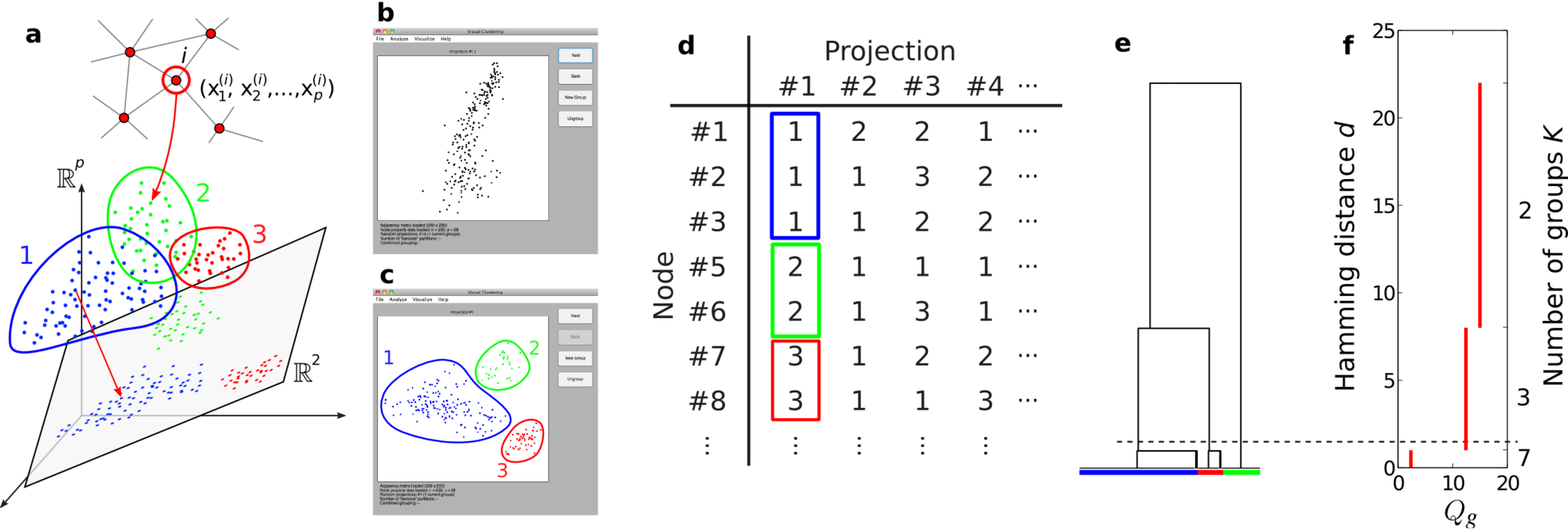}
\caption{\label{fig:method}
{\bf Our visual analytics method.}
 \textbf{a}, The $p$ node properties $x_1^{(i)}, \ldots, x_p^{(i)}$ are computed for each node $i$ in a given network of $n$ nodes. The nodes are then represented as points in $\R^p$, which are projected onto a randomly chosen two-dimensional subspace. 
\textbf{b}, \textbf{c}, Using a graphical interface, the user can either reject the projection (\textbf{b}), which indicates that there is no visible group separation, or 
indicate visible groups (\textbf{c}), which automatically assigns a group index to each node for that particular projection.
\textbf{d}, Repeating this for a given number of random projections, each node $i$ is associated with a group assignment vector $\mathbf{a}^{(i)}$, listing the group indices the user has assigned to node $i$.  We used $30$ projections for all results in this article.
\textbf{e}, Dendrogram obtained by clustering the vectors $\mathbf{a}^{(1)},\dots,\mathbf{a}^{(n)}$.
Cutting the dendrogram at a threshold Hamming distance $d$ produces a grouping for the network.
\textbf{f}, Quality of grouping $\GQ$ as a function of the threshold level $d$.
The appropriate number of groups is determined to be $K=3$ by thresholding at the $\GQ$ drop-off (dashed line).}
\end{figure*}

Our approach overcomes these difficulties using the visual processing ability of a human user as an integral part of the analysis.
The approach is based on visual analytics~\cite{10.1109/MCG.2006.5,Simoff:2008fk},  which is conceptualized as exploratory statistics in which analytical reasoning is facilitated by a visual interactive interface. Humans generally excel automated computer algorithms in visual recognition tasks, such as labeling images~\cite{von2004labeling} and deciphering distorted texts, which forms the basis of spam prevention systems and crowdsourcing for the digitalization of old books~\cite{von2008recaptcha}. We exploit this capability by asking the user to inspect a selection of two-dimensional projections of the node property space for possible separation of nodes into groups.  
Since any projection could potentially reveal good separation of groups, we first consider the result of choosing these projections randomly.  
For two clusters of points with a gap between them in high dimension, the probability can be very small for the clusters to be separable by a straight line in a random two-dimensional projection.
This probability depends strongly on the ``effective dimension'' of the clusters.
For example, if two Gaussian-distributed clusters of 100 points have their centers 6 units apart in the 28-dimensional space, the probability is less than $0.001$ if the variance of the clusters in every direction is one, but increases to about $0.017$ if the variance is reduced by a factor of 10 in all but 10 orthogonal directions. 
We find that the effective dimension is relatively small for the groups discovered in the networks considered here, most of them with dimension less than 12 (out of 28) when defined as the minimum number of principal components required to account for 90\% of the variance within the group.
To further enhance the probability of separating groups, we sample random projections with a systematic bias (see Methods).  
This increases the separation probability for the example of Gaussian clusters above to around $0.68$ for a single projection.  
If the user visually recognizes separation of nodes into groups in a two-dimensional projection, the group assignment is entered through a graphical interactive interface (Fig.~\ref{fig:method}\textbf{a}--\textbf{d}, \href{http://www.nature.com/srep/2011/111109/srep00151/extref/srep00151-s2.mov}{Supplementary Video~S1}).  The integration of the visual component allows the user not only to supervise the process, but also to learn and create intuition from taking part in the process, thus facilitating the search for unanticipated network structures. It also accommodates naturally an ultimate goal of clustering algorithms, which is to reproduce how a human would group a given set of points. 

The chance of capturing a group structure is even further enhanced by the multiplicative effect of using more than one projection.
Indeed, the separation probability in the example above rises from $0.68$ to above $0.999$ with just 7 projections.
In general, for a given number $L$ of random projections, the probability that all of these projections fail to separate a given pair of group decreases to zero exponentially with $L$.
After the user processes a given number $L$ of projections, each node $i$ in the network will be associated with a group assignment vector $\mathbf{a}^{(i)}$  representing the user input (Fig.~\ref{fig:method}\textbf{d}). Since we typically have a large number of distinct assignment  vectors, we aggregate the corresponding nodes into a smaller, more meaningful number of structural groups by single-linkage hierarchical clustering~\cite{jain1999data}.  For this, we use the Hamming distance between the group assignment vectors of different nodes, $\mathbf{a}^{(i)}$ and $\mathbf{a}^{(j)}$, which in this case is the number of projections for which the user has placed those nodes in different groups.  This results in a dendrogram that we can cut at a threshold distance $d$ to obtain a grouping, in which being in different groups indicates that the user has placed these nodes in different groups in at least $d$ out of $L$ projections (Fig.~\ref{fig:method}\textbf{e}; \href{http://www.nature.com/srep/2011/111109/srep00151/extref/srep00151-s2.mov}{Supplementary Video~S1}). To compare the different groupings obtained at different thresholds, we define the \emph{quality of grouping} $\GQ$ by
\begin{equation}\label{eqn:GQ}
\GQ = \frac{1}{\GQbase}\cdot\frac{\langle ||\mathbf{c}_k - \mathbf{c}_\ell|| \rangle_{k,\ell}}{\langle ||\mathbf{x}^{(i)} - \mathbf{c}_{k_i}|| \rangle_{i}},
\end{equation}
where vector $\mathbf{x}^{(i)} = (x_1^{(i)}, x_2^{(i)}, \ldots, x_p^{(i)})^T$  represents node $i$ in the property space, vector $\mathbf{c}_k$ is the center of group $k$, index $k_i$ denotes the group to which node $i$ belongs, and $||\cdot||$ defines the $p$-dimensional Euclidean distance.  The ratio of the two bracketed quantities in Eq.~\eqref{eqn:GQ} measures the average separation distance between groups (the average over all pairs of groups, denoted $\langle\,\cdot\,\rangle_{k,\ell}$) \rule{1pt}{0pt}relative to the spread within individual groups (the average over all nodes, denoted $\langle\,\cdot\,\rangle_{i}$).
This quantity is then normalized by a constant $\GQbase$, chosen to remove a systematic dependence of the quality of grouping on the number of groups $K$ (see Methods).  As one lowers the threshold level, the quality of grouping $\GQ$ tends to drop sharply at a certain level (Fig.~\ref{fig:method}\textbf{f}). To obtain the maximum number of high-quality groups, we suggest choosing the group assignment, as well as the number of groups $K$, at the threshold level just above the largest drop in $\GQ$, which we call the \emph{$\GQ$ drop-off}.

\section*{Results}

\begin{figure}
\includegraphics[width=\columnwidth]{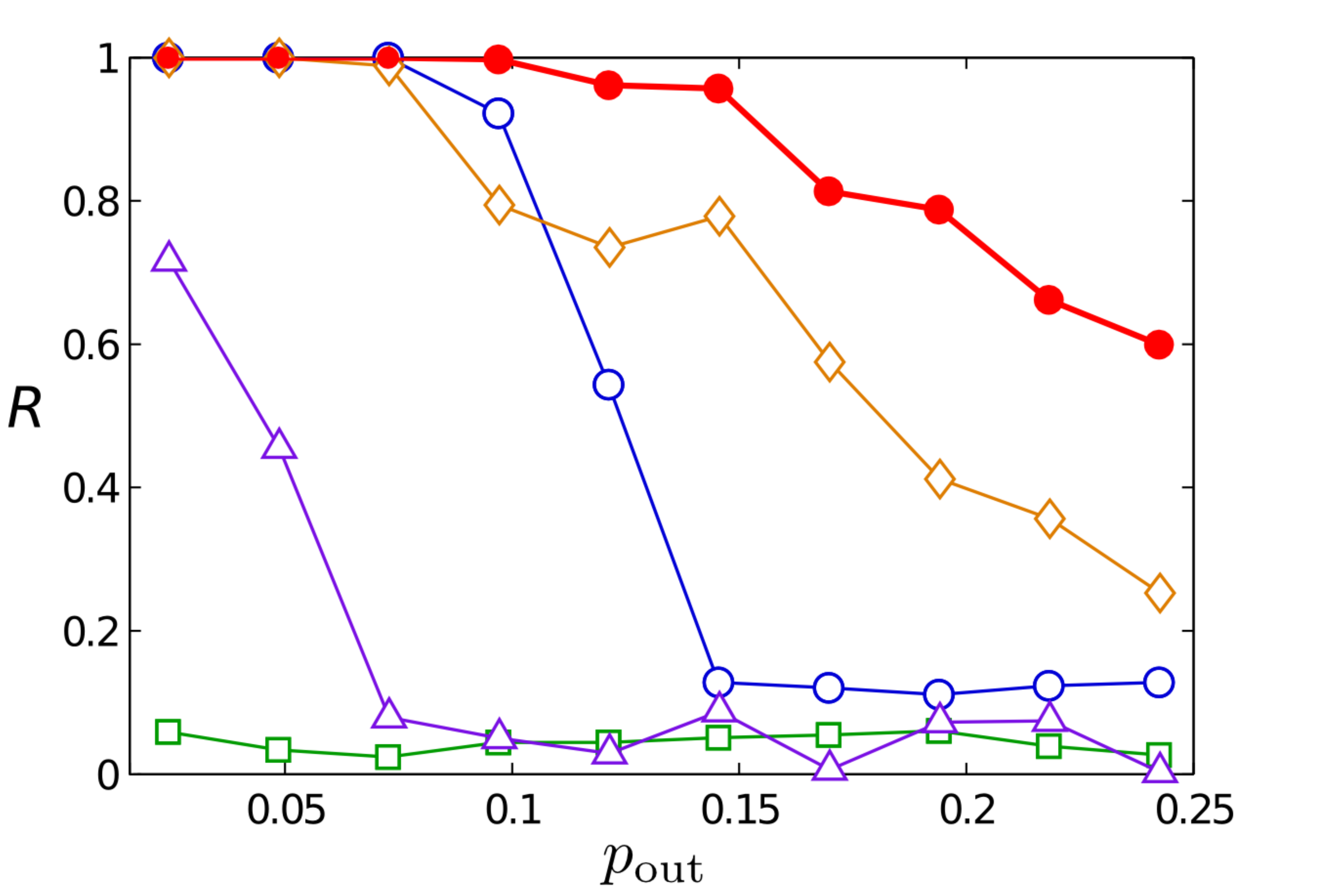}
\caption{\label{fig:benchmarking}
{\bf Performance comparison for detecting density-based communities.}
Using benchmark networks consisting of two groups, we compare the performance of the visual analytics method against 
alternative
methods, measured by the adjusted Rand index~\cite{springerlink:10.1007/BF01908075} $R$ between the computed and the true groupings (see Methods for the details of our benchmarking procedure).  Our method (red filled circles) finds the correct group assignment almost perfectly 
for inter-group connection probability $p_\text{out} \lesssim 0.15$, and performs reasonably well for larger values of
$p_\text{out}$. 
The mixture model method~\cite{Newman:2007rc} (blue open circles) performs well for $p_\text{out} \lesssim 0.10$.
The $K$-means algorithm~\cite{Lloyd.:1982gf} (green open squares) shows
consistently low performance. 
The use of nonlinear kernels, dimensionality reduction based on principal component analysis, and alternative schemes for assigning weights for node properties (see Methods) led to improved performance only for small $p_\text{out}$ values.  One of the largest such improvement is shown here (purple open triangles).
In contrast, replacing the human user with the $K$-means algorithm to process the two-dimensional projections in our method (see Methods) shows significantly better performance (orange open diamonds) than the direct application of the $K$-means variants to the node property space (green open squares and purple open triangles), although still worse than the visual analytics method (red filled circles).  This demonstrates both the effectiveness of the multiple random projection approach and the advantage of the human interactive component over unsupervised algorithms.
Each point in the plot is the average of $R$ computed after removing two
outliers (smallest and largest $R$) from a total of $20$ network realizations.  
The visual analytics method is generally expected to outperform existing methods if the
groups have different internal structures, 
in this case determined by their different degree distributions (see Methods). 
}
\end{figure}

We implemented our {\it visual analytics method} using the selection of $p=28$ node properties listed in Table~\ref{tab:node_properties}, which encompasses important node attributes that capture local information, such as degree and clustering, and others that capture more global information, such as betweenness centrality and Laplacian eigenvectors. In particular, the eigenvectors of the Laplacian  and of the normalized Laplacian allow the detection of communities~\cite{Donetti:2004qy,Fortunato:2010uq,Seary:1995kx,pothen:430,PhysRevE.74.036104} and bipartite or multipartite structures~\cite{chung1997spectral}, respectively, as well as mixtures of these structures, assuring our method the ability to detect group structures defined by link density as special cases. Using this set of properties for the example network of Fig.~\ref{fig:example}, we obtain the dendrogram shown in Fig.~\ref{fig:method}\textbf{e}. The number of groups for this network is found to be $K=3$ at the $\GQ$ drop-off (Fig.~\ref{fig:method}\textbf{f}), which agrees with the group separation visible in the projection shown in Fig.~\ref{fig:example}\textbf{b}. This accurately reflects the fact that the network was synthetically constructed from three distinct structural groups: the first two groups characterized by high ($\ge 65$) and low  ($\le 55$) prescribed degrees, respectively, but connected randomly otherwise, and the third group characterized by higher connection probability with internal nodes ($0.3$) than with external ones ($0.1$). This example illustrates that our method is capable of discovering not only group structures defined by link density, but also more general group structures, even when different types of structures coexist in the same network.
Moreover, as shown in Fig.~\ref{fig:benchmarking} for two-group benchmark networks, the visual analytics method 
is generally expected to 
outperform existing methods if the groups have different internal structures, in this case determined by their different degree distributions (see Methods).

\begin{table}
\caption{\label{tab:node_properties} \bf Node properties used to generate our 
results.}
\begin{ruledtabular}
\begin{tabular}{lp{.86\columnwidth}}
$j$ & $x^{(i)}_j$ ($j$th property of node $i$)\\
\hline
1, 2 & The degree$^a$ of node $i$ and the average degree of the neighbors of node $i$\\
3 & The clustering coefficient$^b$ of node $i$\\
4 & The average shortest path length from node $i$ to all the other nodes\\
5, 6 & The betweenness centrality$^c$ of node $i$ and the average of the same quantity over the neighbors of node $i$\\
7, 8 & The subgraph centrality$^d$ of node $i$ and the average of the same quantity over the neighbors of node $i$\\
9--18 & The $i$th component of the eigenvector associated with the 2nd (smallest 
nonzero) through the 11th eigenvalue of the Laplacian matrix$^e$\\
19--28 & The $i$th components associated with the 10 largest eigenvalues of the 
normalized Laplacian matrix$^f$\\
\end{tabular}
\end{ruledtabular}
\footnotetext[0]{Here we consider only undirected and unweighted networks for 
simplicity. Each quantity is normalized to the unit interval $[0,1]$ before 
applying our analysis, which in this case reduces the node property space to the 
$28$-dimensional unit hypercube.\\
\noindent
$^a$\,The number of links attached to node $i$.\\
$^b$ The fraction of pairs of neighbors of node $i$ that are 
connected.\\
$^c$\,The number of shortest paths passing through node $i$.\\
$^d$\,The weighted sum of the number of closed paths in which node $i
$ participates~\cite{ESTRADA_PRE05}.\\
$^e$\,The Laplacian matrix $L$ is defined by $L_{ij}=-1$ if nodes $i$ 
and $j\neq i$ are connected, $L_{ij}=0$ if they are not connected, and $L_{ij}$ 
equals the degree of node $i$ if $i=j$.\\
$^f$\,The normalized Laplacian matrix is obtained by dividing each 
$L_{ij}$ by the degree of node $i$.}
\end{table}

\begin{figure*}
\includegraphics[width=\textwidth]{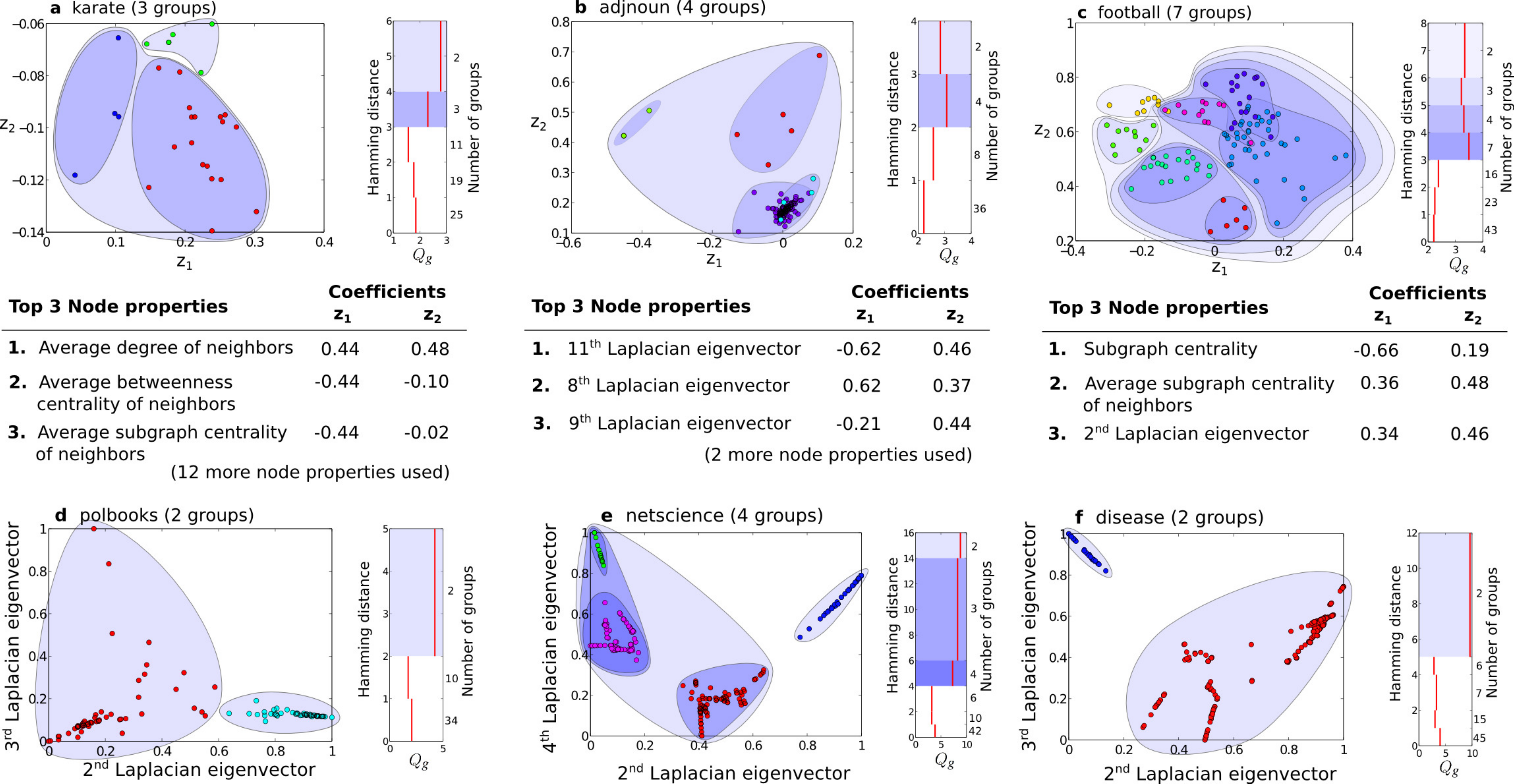}
\caption{\label{fig:results}
{\bf Hierarchical group structures discovered for the networks in Table~\ref{tab:datasets}.}
Each panel shows the network nodes plotted in a two-dimensional projection of the node property space.
For panels \textbf{a}--\textbf{c}, the two coordinate axes, $z_1$ and $z_2$, are linear combinations of a selection of node properties that capture most of the group separation, with corresponding coefficients listed in the table below each plot.
Note that in panels \textbf{b} and \textbf{c} two dimensions are not sufficient to cleanly separate the groups, even though our method resolves this separation by combining multiple two-dimensional projections.
For panels \textbf{d}--\textbf{f}, 
only two properties are necessary to resolve the groups clearly.
The plot on the right of each panel shows the quality of grouping $\GQ$ as a function of the hierarchical level measured by the Hamming distance.
The groupings corresponding to the hierarchical levels in the blue part of this plot are indicated in the projections by the shades of blue.}
\end{figure*}

Figure~\ref{fig:results} shows a visualization of the hierarchy of nested structural groups identified by applying our method to a selection of six real-world networks spanning different sizes and domains (Table~\ref{tab:datasets}).  
To further characterize these groups, we rank the node properties based on a two-dimensional projection in which the discovered groups reveal maximal separation (see Methods). 
We then discard the  low-ranking  properties that have negligible effect on the group separation, keeping only those indicated under each panel.  Surprisingly, while most groups cannot be identified using a single node property,  the node structural groups are completely separated in this plane for four of the networks.  The groups in three of the networks, the polbooks, netscience, and disease networks (Fig.~\ref{fig:results}\textbf{d}-\textbf{f}), are separated using  two eigenvectors of the Laplacian matrix, suggesting that these groups could be similar to density-based communities detected by existing methods~\cite{Girvan:2002fk}; when quantified by the Rand index~\cite{springerlink:10.1007/BF01908075}, however, the similarity appears relatively low (Supplementary Fig.~S2). The  groups in a fourth network, the karate network (Fig.~\ref{fig:results}\textbf{a}), can also be separated in a plane, but this projection requires the use of $15$ properties led by the average degree, average betweenness, and average subgraph centrality~\cite{ESTRADA_PRE05} of neighbors (see Table~\ref{tab:node_properties} notes for the definition).  The groups in the other two networks,  the adjnoun and football networks (Fig.~\ref{fig:results}\textbf{b}-\textbf{c}), are mostly but not completely separated in this two-dimensional representation. 
We emphasize that it is not necessary for all the groups to be separable in a single two-dimensional projection.  In fact, while each such projection may only illuminate part of the hidden group structure (such as the separation between a single group and all the others), the multiplicative effect of integrating information from many random projections is what often reveals the full high-dimensional structure.

\begin{table}
\caption{\label{tab:datasets} \bf Networks analyzed.}
\begin{ruledtabular}
\begin{tabular}{lccccc}
Dataset 							& $n$ 	& $m$	& $K$	& $d_H$ & $p_d$ \\
\hline
{\bfseries karate}			& 34		& 78		& 3		& 3 			& 15 \\
{\bfseries polbooks}		& 105	& 441	& 2		& 2			& 2\\
{\bfseries adjnoun}			& 112	& 425	& 4		& 2			& 5\\
{\bfseries football}		& 115	& 613	& 7		& 3			& 3\\
{\bfseries netscience}	& 379	& 914	& 4		& 4			& 2\\
{\bfseries disease}			& 516	& 1188	& 2		& 5			& 2 
\end{tabular}
\end{ruledtabular}
\footnotetext[0]{We used the following datasets: {\bfseries karate}, the social network of Zachary's Karate club~\cite{1977}; {\bfseries polbooks}, a network of political books~\cite{Krebs}; {\bfseries adjnoun}, a network of English words~\cite{PhysRevE.74.036104}; {\bfseries football}, a network of collegiate American football teams~\cite{Girvan:2002fk}; {\bfseries netscience}, a network of network scientists~\cite{PhysRevE.74.036104}; and {\bfseries disease}, a human disease network~\cite{Goh:2007fk} (see Supplementary Table~SI for a more detailed description of these networks).
Here $n$ is the number of nodes, $m$ is the number of links, $K$ is the number of structural groups at the $\GQ$ drop-off, $d_H$ is the Hamming distance at the $\GQ$ drop-off, and $p_d$ is 
the minimum number of node properties necessary to produce a two-dimensional projection in which most or all of the $K$ discovered groups are visually separated.}
\end{table}

Another remarkable feature of this approach is that, because we do not know in advance which properties define the groups we seek to identify, the visual analytics method simultaneously provides the answer to the question---the number and identity of the structural groups---along with the question itself---the properties that define these groups. Even when these properties are abstract, further analysis can easily reveal the nature of the network's internal organization. For example, consider the karate network, whose nodes are members of a karate club and links are interactions between two members in at least one context external to the club activities. The three structural groups identified in Fig.~\ref{fig:results}\textbf{a} correspond to (1) members who are central to the club and interact with many other members; (2) peripheral members interacting only with very few, but central members; and (3) members forming a community connected to the rest of the network only through one central member (Supplementary Fig.~S3). Incidentally, one of the groups we identify consists of nodes that are connected to those outside the group but to none within the group. 
This social group structure is markedly different from the well-studied eventual split of the club into two clubs~\cite{1977}. 

\begin{figure*}
\includegraphics[width=\textwidth]{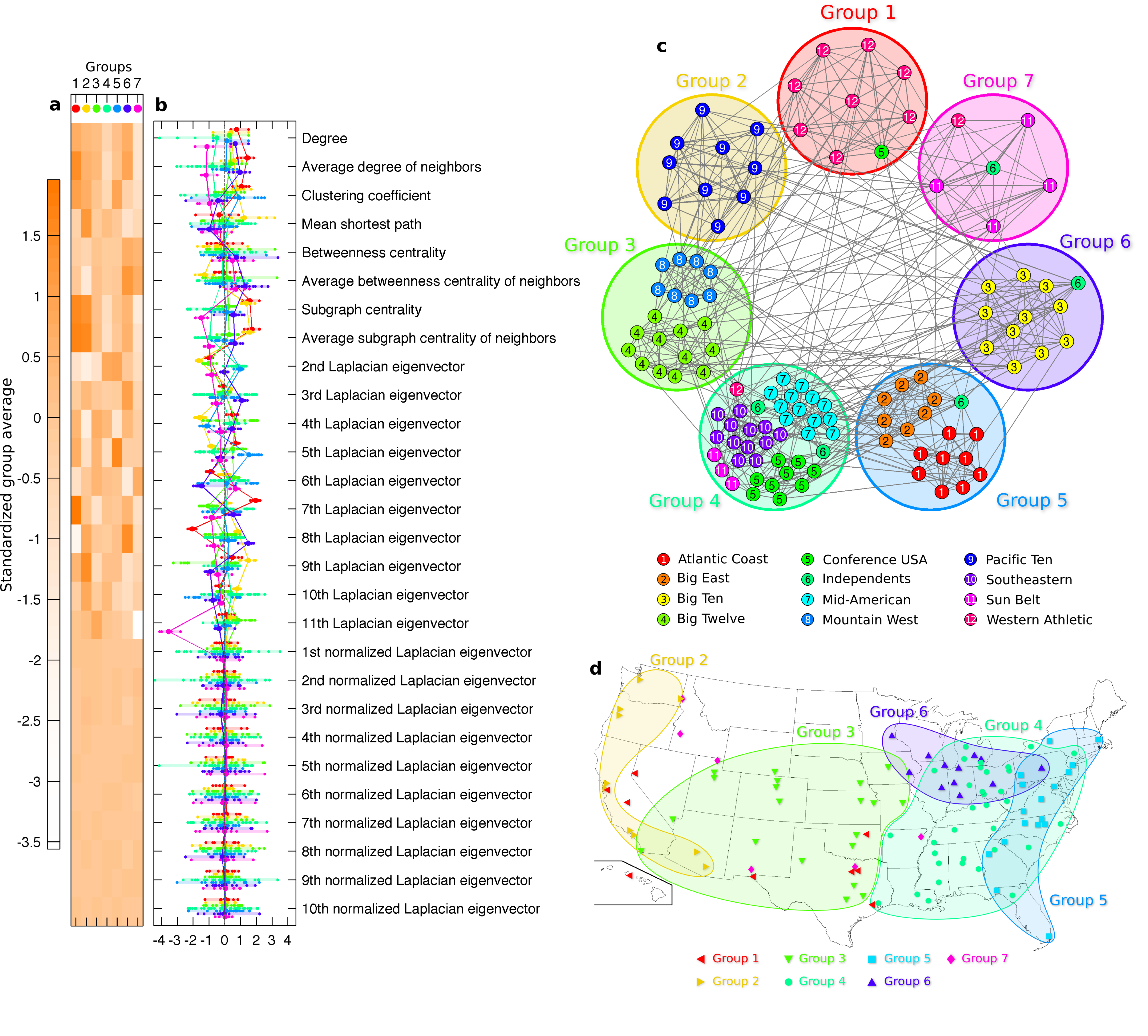}
\caption{\label{fig:football_characterize_groups}
{\bf Characterizing seven structural groups discovered in the football network.}
\textbf{a}, Average node properties of the seven groups.
Rows correspond to node properties and columns to groups (the colored disks at the top).
Using the orange color-scale on the left, each cell shows the average node property of the group, relative to the network average and in units of the network standard deviation.
\textbf{b}, Node property distribution within each group.  
The seven groups in this plot are color-coded as in the disks at the top of panel \textbf{a}.
Small dots indicate the individual values for each node in the network, larger dots connected by lines indicate the group averages, and bars indicate the range of values for each group.
All values are measured relative to the network average and in units of the network standard deviation.   \textbf{c}, Layout of the network with the structural groups indicated by circles, color-coded as in the other panels.  The number and color on a node indicate the college football conference to which the corresponding team belongs, as listed at the bottom of the panel.
\textbf{d}, Geographic distribution of nodes (teams) over the US, color-coded by the structural groups as in panel~\textbf{c}.
The fact that more than one conference is grouped together as groups 3, 4, and 5 can be interpreted in terms of the proximity of the teams' geographic location and its impact on the structure of the network.}
\end{figure*}

As an additional example, consider the football network, where nodes are college American football teams and links indicate matches played in the 2000 season.
Although the teams are organized into 12 conferences (including Independents), our method identifies 7 structural groups (Fig.~\ref{fig:results}\textbf{c}). 
As shown in Figs.~\ref{fig:football_characterize_groups}\textbf{a} and \ref{fig:football_characterize_groups}\textbf{b}, groups 1 and 6 are characterized by the combination of high degrees, high subgraph centrality, and the same characteristics for their neighbors, while these two groups are distinct in clustering coefficient and some Laplacian eigenvectors.
Low degrees and low subgraph centrality, as well as the same characteristics for the neighbors, distinguish groups 4 and 7 from others, while they differ in their clustering coefficient and a few Laplacian eigenvectors.
Group 2 shows similar characteristics as group 1 in terms of subgraph centrality, but the mean shortest path distance is very high and the betweenness centrality of the neighbors is very low, reflecting the peripheral location of these nodes within the network.
Many of the Laplacian eigenvectors contribute to the separation of the groups, which is consistent with the fact that a density-based community structure exists in addition to other group structures.
In particular, groups 3 and 5 are communities that can only be distinguished by the differences in the Laplacian eigenvectors and clustering coefficient. 
Grouping together Big Twelve and Mountain West as well as Atlantic Coast and Big East, but splitting the Independents (Fig.~\ref{fig:football_characterize_groups}\textbf{c}), this group structure captures a higher-level organization of the conferences which is determined by the geographic proximity of the teams (Fig.~\ref{fig:football_characterize_groups}\textbf{d}). Similar geographical manifestation of network communities has recently been observed in the effective boundaries defined by human mobility in the US~\cite{10.1371/journal.pone.0015422} and telecommunications in Great Britain~\cite{10.1371/journal.pone.0014248}.

\section*{Discussion}

The structural groups identified by the visual analytics method are characterized by common network properties.  This provides a foundation for the study of the interplay between form and function in complex networks, as network dynamics (and hence function) is  believed to be strongly influenced by network structure. 
The possibilities are extensive with our approach since the user has complete freedom to choose the set of $p$ node properties. 
Within the wide range of possible structures expressible through these properties, the visual analytics method can help discover a specific group structure of interest and interpret it using a ranking of the node properties.
The approach can be easily adapted 
to identify network structures defined by link rather than node characteristics~\cite{Ahn:2010uq}. Moreover, it can be applied to 
networks whose nodes have quantifiable (but not necessarily structural) properties~\cite{Bianconi:2009fk}, such as age, income and level of education in the case of social networks, 
which remain elusive in  existing network representations.  
Systematic benchmarking using synthetic networks shows that our method has advantages over existing methods in identifying density-based communities with distinct internal structures (red vs.\ blue curve in Fig.~\ref{fig:benchmarking}).
Naturally, existing methods such as the one proposed
in Ref.~\onlinecite{Newman:2007rc} may still be more effective in resolving specific
networks not represented in our benchmarks.
In finding general structural groups beyond density-based communities, the visual analytics method
outperforms the direct application of standard clustering algorithms in the full node property space (Fig.~\ref{fig:example}; Supplementary Fig.~S1; red vs.\ green/purple curve in Fig.~\ref{fig:benchmarking}).
This suggests that our approach also has potential to be an alternative for solving general high-dimensional clustering problems. 
The replacement of the human component in the visual analytics method with a simple heuristics based on $K$-means yields a fully objective unsupervised algorithm, which performs much better than various extensions of $K$-means directly applied to the full node property space (orange vs.\ green/purple curves in Fig.~\ref{fig:benchmarking}).
This highlights the critical role played by the integrative analysis of clustering outputs from multiple projections.
Although the visual analytics method converted to an unsupervised algorithm performs better than standard
unsupervised approaches, the original formulation with the human component is still more effective (red vs.\ orange curve in Fig.~\ref{fig:benchmarking}).
By combining the pattern recognition ability of humans with the processing capability of computers, our visual analytics method can resolve the internal organization of complex networks better than either of them alone.

\section*{Methods}

\noindent{\bf Biased random projections.}
To enhance the probability of resolving group separation, 
we first choose each node property $j$ with probability $r_j$ (while requiring a minimum of four properties) and generate a random projection using those selected properties.
The probability $r_j$ is designed to reflect the relative importance of property $j$ in separating the groups.
We set $r_j := [\bar{v}_j / \max_j (\bar{v}_j)]^\alpha$, where $\bar{v}_j := \sum_k w_k v_{k,j}^2$, and $v_{k,j}$ denotes the $j$th component of the normalized basis vector for the $k$th (out of $N$) one-dimensional projections generated randomly and uniformly.
The weights $w_k$ are given by $w_k := \max_{i}(z_{k,i+1} - z_{k,i})\cdot(i/n)\cdot(1-i/n)$, where $z_{k,1} \le z_{k,2} \le \cdots \le z_{k,n}$ denote the ordered points in the $k$th projection for all $n$ nodes in the network.
The parameter $\alpha$ can be used to adjust the bias strength and was taken to be 2 in all computations.

\medskip\noindent{\bf Controlling for group-size effect in $\GQ$.} 
Since smaller groups naturally tend to have smaller within-group variations, the ratio of the averages in Eq.~\eqref{eqn:GQ} increases with the number of groups $K$, even when the groups are not necessarily better separated.
To correct for this bias, we define $\GQ$ by normalizing the ratio by its expected value $\GQbase$ for randomized groupings with the individual group sizes kept fixed.
We estimated $\GQbase$ by averaging over 100 realizations. 

\medskip\noindent{\bf Two-group benchmark networks.}
For the benchmarking results shown in Fig.~\ref{fig:benchmarking}, we used networks having two groups, constructed as follows.
In the larger group (150 nodes), nodes are connected randomly, with the degree
of each node fixed to a random integer chosen uniformly between 10 and 70.
In the smaller group (50 nodes), node pairs are connected randomly with fixed
probability $p_\text{in}$. Across the two groups, node pairs are connected with
probability $p_\text{out}$. For a given $p_\text{out}$, we choose $p_\text{in}$ to
match the average degree in the smaller group with the average internal degree
in the larger group. The probability $p_\text{out}$ is varied between $0$ (two
completely isolated groups) and $40/150 \approx 0.27$ (no internal links in the
smaller group), with $p_\text{out} = 20/150 \approx 0.13$ corresponding to the
point at which the average internal and external degrees in the smaller group
are equal. 

\medskip\noindent{\bf Benchmarking procedure.}
We used the two-group network described in the subsection above to compare performance of various methods for identifying the groups.
For our visual analytics method, we used the node properties listed
in Table~\ref{tab:node_properties} and generated 30 biased random projections. The threshold
level for the resulting dendrogram was selected so as to produce two groups.
In a few cases where a two-group threshold does not exist, we selected the threshold
that results in the smallest possible number of groups above two.
For the mixture model method~\cite{Newman:2007rc}, the number of groups was set to $K=2$.
For $K$-means~\cite{Lloyd.:1982gf}, the algorithm was applied directly to the node
property space with $K=2$.
For completeness, we also examined the performance of $K$-means using all possible combinations of choices for (i) kernel~\cite{Scholkopf:1998lr} (linear, polynomial, Gaussian, or sigmoid); (ii) dimensionality reduction (projecting the data points in the node property space onto the 2, 5, 10, 15, or 20 leading principal components, or no reduction); and (iii) normalization (scaling each node property to have zero mean and unit variance, normalizing each property to the unit interval $[0,1]$, or no normalization).  Scaling for zero mean and unit variance is equivalent to weighing each node property equally when measuring distances in the node property space, while normalizing to the unit interval ensures that all the node properties are distributed in the same range.  For the unsupervised variant of our visual analytics method, the human user was replaced by the (linear) $K$-means algorithm with $K=1,2,\ldots,10$ to analyze each two-dimensional projection, with an optimal choice of $K$ determined by the gap statistic~\cite{Tibshirani:2001lr}, which is defined based on a characteristic signature in the $K$-dependence of the within-group variation.
The performance of each method was measured by the adjusted Rand index $R$ between the computed and the true groupings (see the subsection below for definition).

\medskip\noindent{\bf Rand index.}
This index measures the similarity between two ways of grouping a given set of discrete objects, possibly into different numbers of groups.
For a given pair of groupings of network nodes, the adjusted Rand index $R$ is defined as the normalized fraction of node pairs that are either classified in the same group in both groupings or classified in different groups in both groupings~\cite{springerlink:10.1007/BF01908075}.
The normalization implies that $R = 1$ for identical groupings and $R \approx 0$ for a pair of random groupings.

\medskip\noindent{\bf Ranking node properties.} 
For a given node grouping, we seek a two-dimensional projection that maximizes 
$\langle n_k||\mathbf{c}_k - \mathbf{c}||^2 \rangle_{k} / \langle ||\mathbf{x}^{(i)} - \mathbf{c}_{k_i}||^2 \rangle_{i},$
a group separation measure similar to that in Eq.~\eqref{eqn:GQ} but computed for the projected points after the groups have been identified.
Here $n_k$ denotes the number of nodes in group $k$, and $\mathbf{c}$ denotes the center of all the data points.
Such a projection plane can be efficiently found by a spectral method~\cite{park2007fast} based on the QR decomposition.
The node properties are then ranked in the order of increasing angle between their coordinate axes and the projection plane.

\section*{Software} A version of the visual analytics software that implements our method for all the networks discussed in this article is available at \url{http://purl.oclc.org/net/find_structural_groups}

\begin{acknowledgments}
This work was supported by NSF DMS/FODAVA Grant No. 0808860.
\end{acknowledgments}

\section*{Author Contributions}
T.N. and A.E.M. designed the research, performed the
research, and wrote the manuscript.


\clearpage
\onecolumngrid

\setcounter{page}{1}
\renewcommand{\thepage}{{S-\arabic{page}}}

\setcounter{figure}{0}
\renewcommand{\figurename}{{Supplementary Figure}}
\renewcommand{\thefigure}{{S\arabic{figure}}}

\begin{center}
\vspace{50mm}
{\LARGE\bf Supplementary Information}\\[15mm]
\end{center}

\begin{figure*}[h!]
\begin{center}
\includegraphics[width=0.95\textwidth]{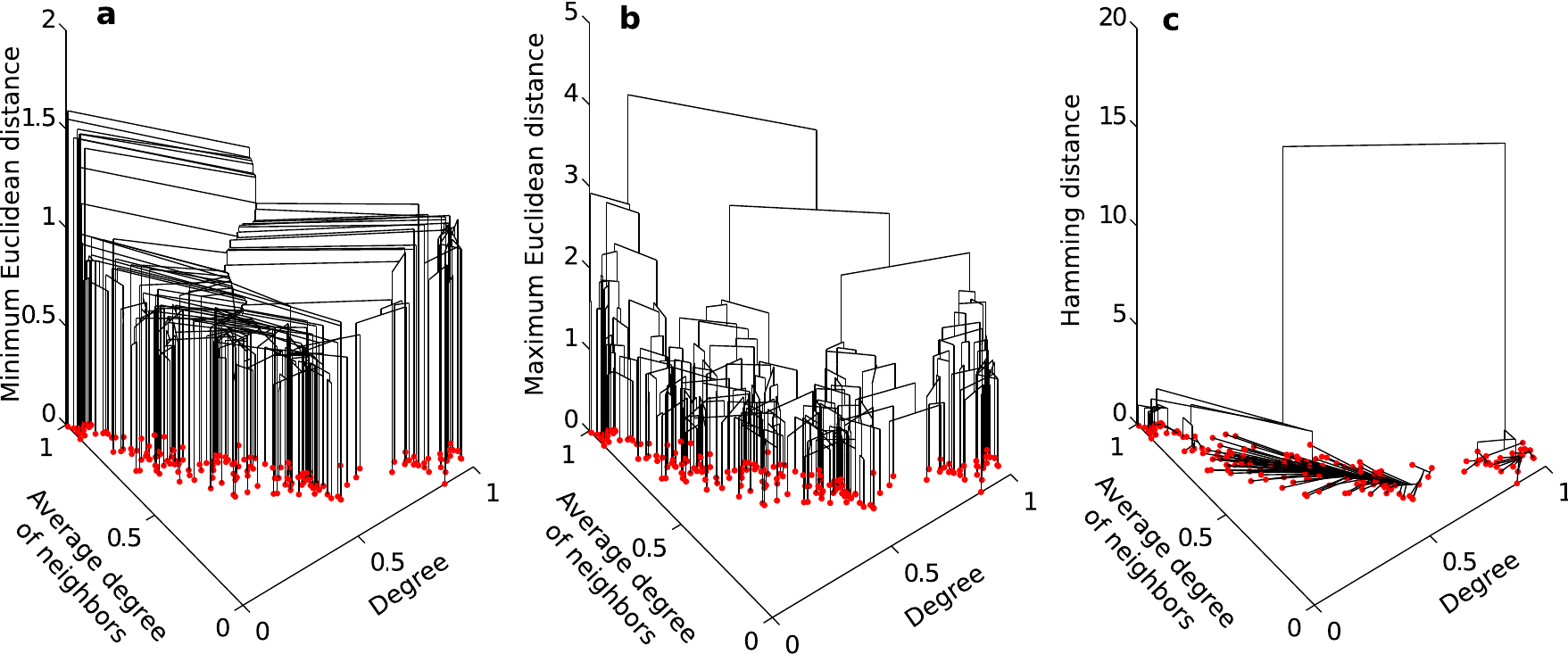}
\vspace{5mm}
\caption{
\label{fig:hierarchical}
{\bf Comparing hierarchical clustering methods.}
\textbf{a}, \textbf{b}, Dendrograms obtained by applying the single-linkage (\textbf{a}) and complete-linkage (\textbf{b}) clustering algorithms$^{32}$
to the full $28$-dimensional node property space.
We use an example network with two structural groups defined by their degree distributions, which we  construct by randomly connecting nodes with prescribed degrees.
A group of 150 low-degree nodes (degree between 10 and 50) and a group of 30 high-degree nodes (degree between 70 and 90) form a bimodal degree distribution, bridged by 5 nodes with degree between 50 and 70.  All node degrees are chosen uniformly at random from the corresponding intervals.
The nodes (red dots) are plotted in the two-dimensional projection using the degree and the average degree of neighbors, both normalized to the unit interval.
For these two dendrograms, the distance between two nodes is measured by the Euclidean distance in the full node property space.
In the single-linkage dendrogram (\textbf{a}) the height at which two groups join is the minimum distance between all pairs of nodes from the two groups, while in the complete-linkage dendrogram (\textbf{b}) it is the maximum distance between node pairs.
\textbf{c}, Dendrogram obtained by our visual analytics method.
In applying single-linkage clustering to obtain the dendrogram, the measure used for distance between nodes $i$ and $j$ is 
the Hamming distance between the user input vectors $\mathbf{a}^{(i)}$ and $\mathbf{a}^{(j)}$ (which can be defined in this case, but not in the other two cases).
Two or more groups joined in this dendrogram at height $d$ implies that any pair of nodes taken from two of these groups has been separated visually by the user in at least $d$ different projections.
We see that thresholding at a fixed level in this dendrogram would clearly produce the correct splitting of nodes into the two structural groups, while the dendrograms in the other two panels do not capture the group structure.
This comes from the difficulty in recognizing groups with wide distribution of points, also suffered by the $K$-means algorithm (see Fig.~1\textbf{d}).
}
\end{center}
\end{figure*}

\begin{figure*}
\begin{center}
\includegraphics[width=4in]{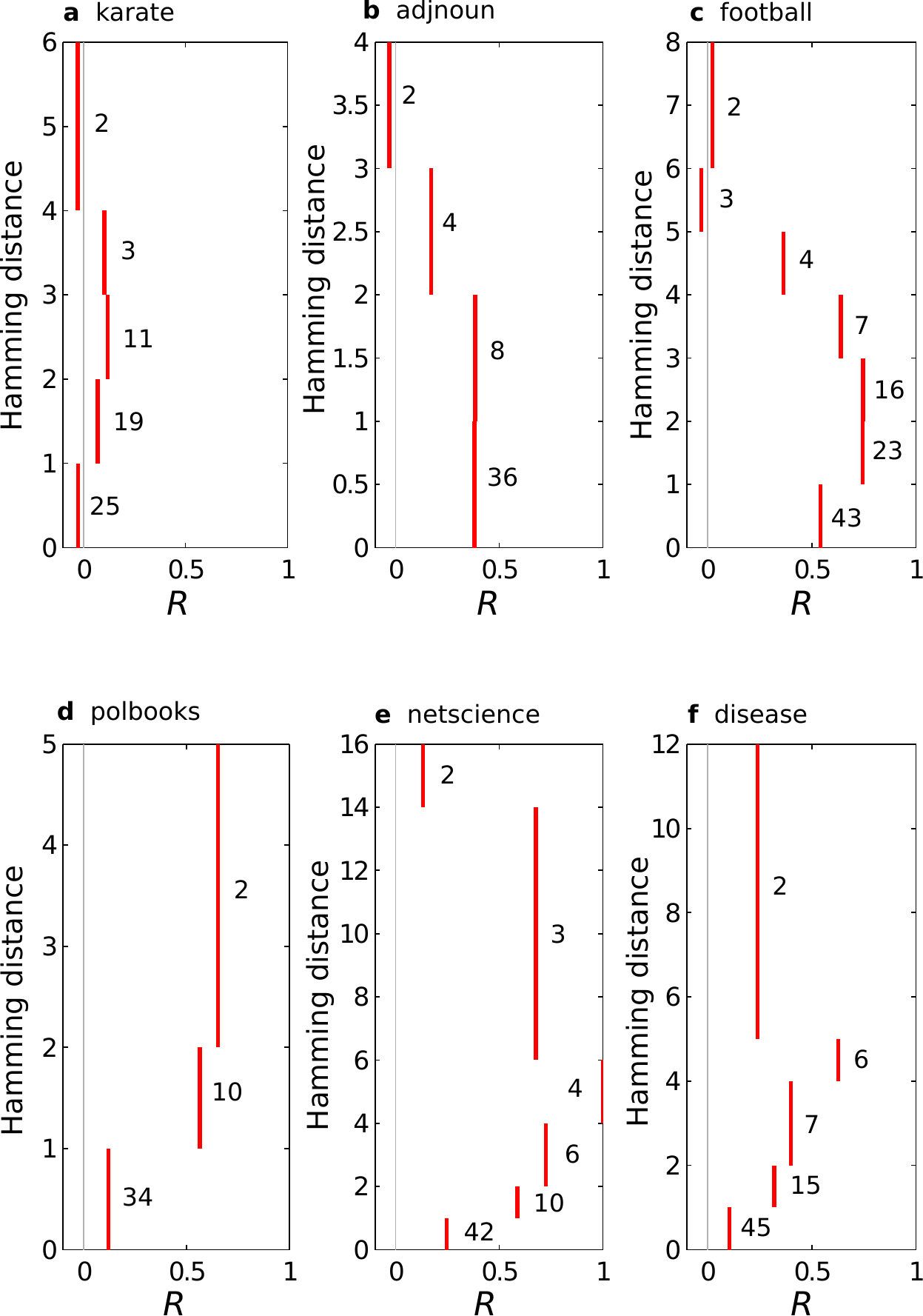}
\end{center}
\caption{{\bf Comparing discovered groups with communities found by a traditional method.}
For each network in Table~II, we compare the grouping obtained at a fixed hierarchical level in our method with the division into the same number of groups that is found by the method of Ref.~2, 
which is based on link betweenness centrality.
The similarity of the two groupings is measured by the adjusted Rand index $R$ (defined in Methods).
Each panel shows $R$ as a function of the hierarchical level measured by the Hamming distance in our method, with the corresponding number of groups indicated in the plots.
The $R$ values significantly lower than unity are observed at most hierarchical levels for each network, confirming that the group structures we discovered are indeed different from the traditional community structures.}
\end{figure*}

\begin{figure*}[h!]
\begin{center}
\includegraphics[width=0.95\textwidth]{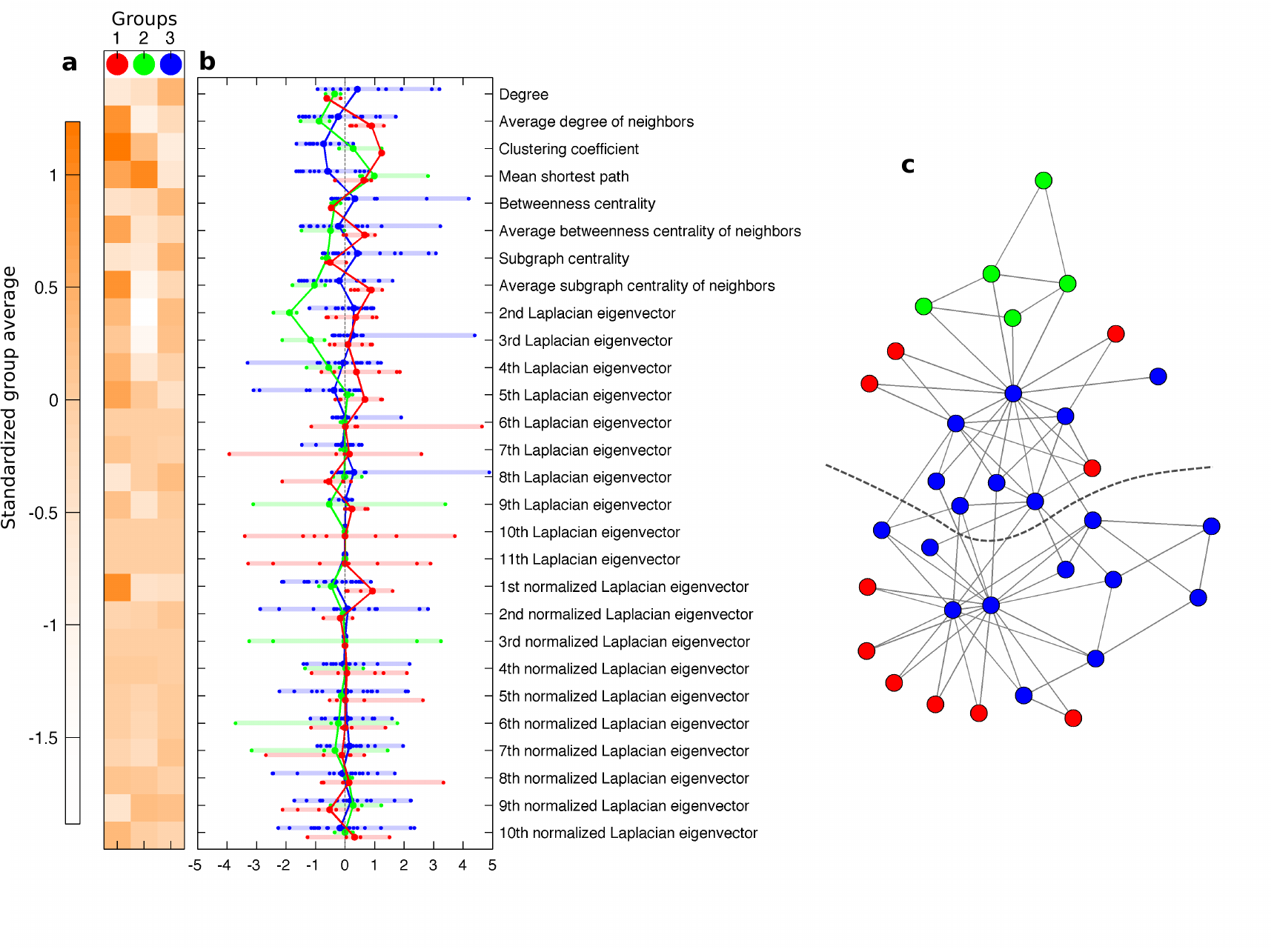}
\vspace{5mm}
\caption{
\label{fig:karate_characterize_3_groups}
{\bf Characterizing three structural groups discovered in the karate network.}
\textbf{a}, Average node properties of the three groups.
Rows correspond to node properties and columns to groups (red, green, and blue disks at the top).
Using the orange color-scale on the left, each cell shows the average node property of the group, relative to the network average and in units of the network standard deviation.
\textbf{b}, Node property distribution within each group.  
The three groups in this plot are color-coded as in the disks at the top of panel \textbf{a}.
Small dots indicate the individual values for each node in the network, larger dots connected by lines indicate the group averages, and bars indicate the range of values for each group.
All values are measured relative to the network average and in units of the network standard deviation.
\textbf{c}, Layout of the network with groups color-coded as in the other panels.  
The dashed curve indicates the eventual split of the karate club into two different clubs, as documented in Ref.~40.
It is clear that the group structure discovered by our visual analytics approach is distinct from the club split.  Group 1 (red) is characterized by low degree, clustering coefficient of one, and large values of the components of the first normalized Laplacian eigenvector, as well as by being connected to high-degree nodes with high betweenness and subgraph centrality (the first column of panel \textbf{a} and red plots in panel \textbf{b}).
Group 2 (green) is characterized by very low values of the components of the second Laplacian eigenvector, which is indicative of a traditional community structure.  This is reflected in panel \textbf{c}, where green nodes form a cluster at the top. 
High values of mean shortest path distance to other nodes for nodes in Groups 1 and 2 implies that these nodes sit in peripheral locations within the network, which is confirmed in the network layout in panel \textbf{c}. 
Group 3 (blue) forms the core of the network and includes all the high-degree, high-centrality nodes, which is reflected in the high group average in these quantities.
This example network illustrates that our method identifies groups of nodes with common structural properties that do not even need to be connected within each group.}
\end{center}
\end{figure*}

\setcounter{figure}{0}
\renewcommand{\figurename}{{Supplementary Video}}
\renewcommand{\thefigure}{{S\arabic{figure}}}
\begin{figure*}
\begin{center}
\href{http://www.nature.com/srep/2011/111109/srep00151/extref/srep00151-s2.mov}{
\includegraphics[width=0.8\textwidth]{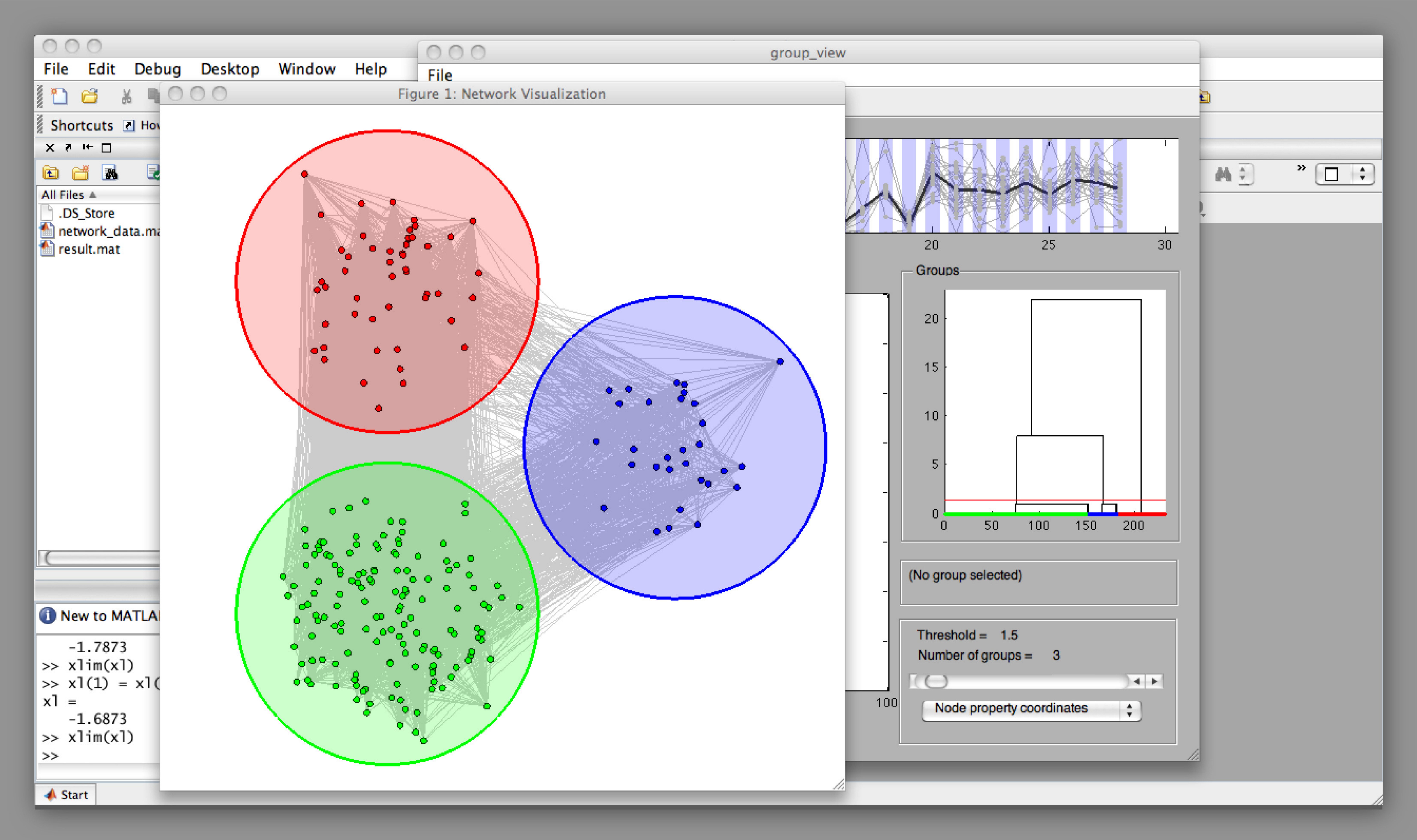}
}\\
{\small Click on the image to see the video on the {\it Scientific Reports} website (MOV Format, Length 4:47).}\\
\end{center}
\vspace{5mm}
\caption{{\bf Visual analytics software for discovering structural groups in networks.}  The network shown in Fig.~1 is used to demonstrate how an implementation of the method introduced in this article discovers the three groups of nodes characterized by the degree and the average degree of neighbors.  The default set of node properties is the same as in Table~I, and we use 30 random projections with the bias described in Methods to enhance the probability that the user sees group separation.
Only a selection of these projections are actually shown in the movie in order to keep the presentation short.}
\end{figure*}

\clearpage
\setcounter{table}{0}
\renewcommand{\tablename}{{Supplementary Table}}
\renewcommand{\thetable}{{S\Roman{table}}}

\begin{table*}
\caption{\bf Description of networks analyzed.}
\begin{center}
\begin{tabular*}{\hsize}{@{\extracolsep{\fill}}lrp{4.7in}c}
\hline
	\multicolumn{2}{l}{Dataset}	& Description & Reference \\
\hline
\multicolumn{2}{l}{\textbf{karate}} & Social network of a university-based karate club & 40 \\
										& Node:	& A club member \\
										& Link:	& Interaction between the two members in at least one context outside the club activity \\
\multicolumn{2}{l}{\textbf{polbooks}} & Network of books on American politics bought from Amazon.com & 49\\
										& Node:	& A book \\
										& Link:	& Frequent purchase of the two books together by the same buyer \\
\multicolumn{2}{l}{\textbf{adjnoun}} & Network of nouns and adjectives appearing in a novel (\textit{David Copperfield} by Charles Dickens)& 36 \\
										& Node:	& A noun or adjective \\
										& Link:	& Appearance of the two words adjacent to each other in the book \\
\multicolumn{2}{l}{\textbf{football}} & Network of collegiate American football teams in the US & 2\\
										& Node:	& A football team\\
										& Link:	& The fact that one or more games were played between the two teams in the 2000 regular season \\
\multicolumn{2}{l}{\textbf{netscience}} & Largest connected component of the network of scientists who have published papers on network science & 36 \\
										& Node:	& A scientist \\
										& Link:	& Coauthorship between the two scientists  \\
\multicolumn{2}{l}{\textbf{disease}} & Largest connected component of the network of known human genetic disorders & 50\\
										& Node:	& A genetic disorder \\
										& Link:	& Existence of a common gene whose mutation is associated with both disorders \\
\hline

\end{tabular*}
\end{center}
\end{table*}


\begin{thebibliography}{10}
\expandafter\ifx\csname url\endcsname\relax
  \def\url#1{\texttt{#1}}\fi
\expandafter\ifx\csname urlprefix\endcsname\relax\def\urlprefix{URL }\fi
\providecommand{\bibinfo}[2]{#2}
\providecommand{\eprint}[2][]{\url{#2}}

\bibitem{Strogatz:2001il}
Strogatz, S. H. 
Exploring complex networks.
\textit{Nature} \textbf{410}, 268--276 (2001).

\bibitem{Girvan:2002fk}
\bibinfo{author}{Girvan, M.} \& \bibinfo{author}{Newman, M. E. J.}
\newblock \bibinfo{title}{Community structure in social and biological
  networks}.
\newblock \emph{\bibinfo{journal}{Proc. Natl. Acad. Sci. USA}}
  {\bf \bibinfo{volume}{99}}, \bibinfo{pages}{7821--7826} (2002).

\bibitem{Guimera:2004gf}
Guimera, R., Sales-Pardo, M. \& Amaral, L.
\newblock Modularity from fluctuations in random graphs and complex networks.
\newblock {\em Phys. Rev. E} {\bf 70}, 025101 (2004).

\bibitem{Radicchi:2004ve}
Radicchi, F., Castellano, C., Cecconi, F., Loreto, V. \& Parisi, D.
\newblock Defining and identifying communities in networks.
\newblock \emph{\bibinfo{journal}{Proc. Natl. Acad. Sci. USA}} {\bf 101}, 2658--2663 (2004).

\bibitem{Palla:2005fk}
\bibinfo{author}{Palla, G.}, \bibinfo{author}{Derenyi, I.},
  \bibinfo{author}{Farkas, I.} \& \bibinfo{author}{Vicsek, T.} 
\newblock \bibinfo{title}{Uncovering the overlapping community structure of
  complex networks in nature and society}.
\newblock \emph{\bibinfo{journal}{Nature}} {\bf\bibinfo{volume}{435}}, \bibinfo{pages}{814--818} (\bibinfo{year}{2005}).

\bibitem{PhysRevE.74.016110}
Reichardt, J. \& Bornholdt, S.
\newblock Statistical mechanics of community detection.
\newblock {\em Phys. Rev. E} {\bf 74}, 016110 (2006).

\bibitem{Danon:2006zr}
Danon, L., Diaz-Guilera, A. \& Arenas, A.
\newblock The effect of size heterogeneity on community identification in
  complex networks.
\newblock {\em J. Stat. Mech.-Theory E.} {\bf 2006}, P11010 (2006).

\bibitem{Fortunato:2007fr}
Fortunato, S. \& Barthelemy, M.
\newblock Resolution limit in community detection.
\newblock \emph{\bibinfo{journal}{Proc. Natl. Acad. Sci. USA}} {\bf 104}, 36--41 (2007).

\bibitem{Chauhan:2009uq}
Chauhan, S., Girvan, M. \& Ott, E.
\newblock Spectral properties of networks with community structure.
\newblock {\em Phys. Rev. E} {\bf 80}, 056114 (2009).

\bibitem{Porter:2009ht}
\bibinfo{author}{Porter, M. A.}, \bibinfo{author}{Onnela, J. P.} \&
  \bibinfo{author}{Mucha, P. J.}
\newblock \bibinfo{title}{Communities in networks}.
\newblock \emph{\bibinfo{journal}{Notices Amer. Math. Soc.}}
  {\bf\bibinfo{volume}{56}}, \bibinfo{pages}{1082--1097} (\bibinfo{year}{2009}).

\bibitem{Chen2010278}
\bibinfo{author}{Chen, P.} \& \bibinfo{author}{Redner, S.}
\newblock \bibinfo{title}{Community structure of the physical review citation
  network}.
\newblock \emph{\bibinfo{journal}{J. Informetr.}} {\bf\bibinfo{volume}{4}}, \bibinfo{pages}{278--290} (\bibinfo{year}{2010}).

\bibitem{Fortunato:2010uq}
\bibinfo{author}{Fortunato, S.}
\newblock \bibinfo{title}{Community detection in graphs}.
\newblock \emph{\bibinfo{journal}{Phys. Rep.}} {\bf\bibinfo{volume}{486}}, \bibinfo{pages}{75--174} (\bibinfo{year}{2010}).

\bibitem{Mucha:2010fk}
Mucha, P. J., Richardson, T., Macon, K., Porter, M. A. \& Onnela, J. P.
\newblock Community structure in time-dependent, multiscale, and multiplex
  networks.
\newblock {\em Science} {\bf 328}, 876--878 (2010).

\bibitem{Newman:2007rc}
\bibinfo{author}{Newman, M. E. J.} \& \bibinfo{author}{Leicht, E. A.}
\newblock \bibinfo{title}{Mixture models and exploratory analysis in networks}.
\newblock \emph{\bibinfo{journal}{Proc. Natl. Acad. Sci. USA}}
 {\bf\bibinfo{volume}{104}}, \bibinfo{pages}{9564--9569} (\bibinfo{year}{2007}).

\bibitem{ravasz2002hierarchical}
\bibinfo{author}{Ravasz, E.}, \bibinfo{author}{Somera, A.},
  \bibinfo{author}{Mongru, D.}, \bibinfo{author}{Oltvai, Z.} \&
  \bibinfo{author}{Barab{\'a}si, A.}
\newblock \bibinfo{title}{{Hierarchical organization of modularity in metabolic
  networks}}.
\newblock \emph{\bibinfo{journal}{Science}} {\bf\bibinfo{volume}{297}}, \bibinfo{pages}{1551--1555} (\bibinfo{year}{2002}).

\bibitem{PhysRevE.75.036105}
\bibinfo{author}{Sreenivasan, S.}, \bibinfo{author}{Cohen, R.},
  \bibinfo{author}{L\'opez, E.}, \bibinfo{author}{Toroczkai, Z.} \&
  \bibinfo{author}{Stanley, H. E.}
\newblock \bibinfo{title}{Structural bottlenecks for communication in
  networks}.
\newblock \emph{\bibinfo{journal}{Phys. Rev. E}} {\bf\bibinfo{volume}{75}}, \bibinfo{pages}{036105} (\bibinfo{year}{2007}).

\bibitem{ISI:000286468600016}
Costa, L.~da~F., Villas~Boas, P.~R., Silva, F.~N. \& Rodrigues, F.~A. 
\newblock {A pattern recognition approach to complex networks}.
\newblock {\em J. Stat. Mech.} {\bf 2010}, P11015 (2010).

\bibitem{Bagrow:2005mz}
Bagrow, J. \& Bollt, E.
\newblock Local method for detecting communities.
\newblock {\it Phys. Rev. E} {\bf 72}, 046108 (2005).

\bibitem{Clauset:2005ly}
Clauset, A.
\newblock Finding local community structure in networks.
\newblock {\it Phys. Rev. E} {\bf 72}, 026132 (2005).

\bibitem{Raghavan:2007rt}
Raghavan, U.~N., Albert, R. \& Kumara, S.
\newblock Near linear time algorithm to detect community structures in
  large-scale networks.
\newblock {\it Phys. Rev. E} {\bf 76}, 036106 (2007).

\bibitem{Rosvall:2007ys}
Rosvall, M. \& Bergstrom, C.~T.
\newblock An information-theoretic framework for resolving community structure
  in complex networks.
\newblock \emph{\bibinfo{journal}{Proc. Natl. Acad. Sci. USA}} {\bf 104}, 7327--7331 (2007).

\bibitem{PhysRevLett.100.258701}
Hofman, J.~M. \& Wiggins, C.~H.
\newblock Bayesian approach to network modularity.
\newblock {\it Phys. Rev. Lett.} {\bf 100}, 258701 (2008).

\bibitem{Kovacs:2010qy}
Kovacs, I., Palotai, R., Szalay, M. \& Csermely, P.
\newblock Community landscapes: An integrative approach to determine
  overlapping network module hierarchy, identify key nodes and predict network
  dynamics.
\newblock {\em PLoS ONE} {\bf 5}, e12528 (2010).

\bibitem{Estrada:2011fj}
Estrada, E.
\newblock Community detection based on network communicability.
\newblock {\em Chaos} {\bf 21}, 016103 (2011).

\bibitem{Lancichinetti:2011yq}
Lancichinetti, A., Radicchi, F., Ramasco, J.~J. \& Fortunato, S.
\newblock Finding statistically significant communities in networks.
\newblock {\em PLoS ONE} {\bf 6}, e18961 (2011).

\bibitem{Psorakis:2011vn}
Psorakis, I., Roberts, S., Ebden, M. \& Sheldon, B.
\newblock Overlapping community detection using bayesian non-negative matrix
  factorization.
\newblock {\it Phys. Rev. E} {\bf 83}, 066114 (2011).

\bibitem{Wen:2011kx}
Wen, H., Leicht, E.~A. \& D'Souza, R.~M.
\newblock Improving community detection in networks by targeted node removal.
\newblock {\it Phys. Rev. E} {\bf 83}, 016114 (2011).

\bibitem{10.1109/MCG.2006.5}
\bibinfo{author}{Thomas, J. J.} \& \bibinfo{author}{Cook, K. A.}
\newblock \bibinfo{title}{A visual analytics agenda}.
\newblock \emph{\bibinfo{journal}{IEEE Comput. Graph.}}
 {\bf\bibinfo{volume}{26}}, \bibinfo{pages}{10--13} (\bibinfo{year}{2006}).

\bibitem{Simoff:2008fk}
Keim, D., Mansmann, F., Schneidewind, J., Thomas, J., \& Ziegler, H. Visual Analytics: Scope and Challenges, in {\it Visual Data Mining}, eds. Simoff, S. J., Bšhlen, M. H. \& Mazeika, A., Vol. 4404 of {\it Lec. Notes Comput. Sc.}, 76-90 (Springer, Berlin/Heidelberg, 2008).
  
  \bibitem{von2004labeling}
\bibinfo{author}{von~Ahn, L.} \& \bibinfo{author}{Dabbish, L.} 
\newblock \bibinfo{title}{{Labeling images with a computer game}},
in \emph{\bibinfo{booktitle}{Proceedings of the SIGCHI conference on
  human factors in computing systems}}, \bibinfo{pages}{319--326}
  (\bibinfo{organization}{ACM}, \bibinfo{year}{2004}).

\bibitem{von2008recaptcha}
\bibinfo{author}{von~Ahn, L.}, \bibinfo{author}{Maurer, B.},
  \bibinfo{author}{McMillen, C.}, \bibinfo{author}{Abraham, D.} \&
  \bibinfo{author}{Blum, M.} 
  \newblock \bibinfo{title}{{reCAPTCHA: Human-based character recognition via web
  security measures}}.
\newblock \emph{\bibinfo{journal}{Science}} {\bf\bibinfo{volume}{321}}, \bibinfo{pages}{1465--1468} (\bibinfo{year}{2008}).

\bibitem{jain1999data}
\bibinfo{author}{Jain, A.}, \bibinfo{author}{Murty, M.} \&
  \bibinfo{author}{Flynn, P.}
\newblock \bibinfo{title}{{Data clustering: A review}}.
\newblock \emph{\bibinfo{journal}{ACM Comput. Surv.}}
 {\bf\bibinfo{volume}{31}}, \bibinfo{pages}{264--323} (\bibinfo{year}{1999}).
  
\bibitem{Donetti:2004qy}
Donetti, L. \& Mu{\~n}oz, M.~A.
\newblock Detecting network communities: a new systematic and efficient algorithm.
\newblock {\em J. Stat. Mech.} {\bf 2004}, P10012 (2004).

  \bibitem{Seary:1995kx}
\bibinfo{author}{Seary, A. J.} \& \bibinfo{author}{Richards, W. D.}
\newblock \bibinfo{title}{Partitioning networks by eigenvectors},
\newblock in \emph{\bibinfo{booktitle}{Proceedings of the International
  Conference on Social Networks}}, Vol.~\bibinfo{volume}{1},
  \bibinfo{pages}{47--58} (\bibinfo{year}{1995}).

\bibitem{pothen:430}
\bibinfo{author}{Pothen, A.}, \bibinfo{author}{Simon, H. D.} \&
  \bibinfo{author}{Liou, K. P.}
\newblock \bibinfo{title}{Partitioning sparse matrices with eigenvectors of
  graphs}.
\newblock \emph{\bibinfo{journal}{SIAM J. Matrix Anal. Appl.}}
 {\bf\bibinfo{volume}{11}}, \bibinfo{pages}{430--452} (\bibinfo{year}{1990}).

\bibitem{PhysRevE.74.036104}
\bibinfo{author}{Newman, M. E. J.}
\newblock \bibinfo{title}{Finding community structure in networks using the
  eigenvectors of matrices}.
\newblock \emph{\bibinfo{journal}{Phys. Rev. E}} {\bf\bibinfo{volume}{74}}, \bibinfo{pages}{036104} (\bibinfo{year}{2006}).

\bibitem{chung1997spectral}
\bibinfo{author}{Chung, F.} 
\newblock \emph{\bibinfo{title}{{Spectral Graph Theory}}}
  (\bibinfo{publisher}{American Mathematical Society},
  \bibinfo{address}{Province}, \bibinfo{year}{1997}).

\bibitem{springerlink:10.1007/BF01908075}
\bibinfo{author}{Hubert, L.} \& \bibinfo{author}{Arabie, P.} 
\newblock \bibinfo{title}{Comparing partitions}.
\newblock \emph{\bibinfo{journal}{J. Classif.}} {\bf\bibinfo{volume}{2}}, \bibinfo{pages}{193--218} (\bibinfo{year}{1985}).

\bibitem{ESTRADA_PRE05}
\bibinfo{author}{Estrada, E.} \& \bibinfo{author}{Rodr\'iguez-Vel\'azquez, J. A.}
\newblock \bibinfo{title}{Subgraph centrality in complex networks}.
\newblock \emph{\bibinfo{journal}{Phys. Rev. E}} {\bf\bibinfo{volume}{71}}, \bibinfo{pages}{056103} (\bibinfo{year}{2005}).
  
\bibitem{1977}
\bibinfo{author}{Zachary, W. W.}
\newblock \bibinfo{title}{An information flow model for conflict and fission in
  small groups}.
\newblock \emph{\bibinfo{journal}{J. Anthro. Res.}}
 {\bf\bibinfo{volume}{33}}, \bibinfo{pages}{452--473} (\bibinfo{year}{1977}).

\bibitem{10.1371/journal.pone.0015422}
\bibinfo{author}{Thiemann, C.}, \bibinfo{author}{Theis, F.},
  \bibinfo{author}{Grady, D.}, \bibinfo{author}{Brune, R.} \&
  \bibinfo{author}{Brockmann, D.}
\newblock \bibinfo{title}{The structure of borders in a small world}.
\newblock \emph{\bibinfo{journal}{PLoS ONE}} {\bf\bibinfo{volume}{5}}, \bibinfo{pages}{e15422} (\bibinfo{year}{2010}).

\bibitem{10.1371/journal.pone.0014248}
\bibinfo{author}{Ratti, C.}, \bibinfo{author}{Sobolevsky, S.}, 
\bibinfo{author}{Calabrese, F.}, \bibinfo{author}{Andris, C.},
\bibinfo{author}{Reades, J.}, \bibinfo{author}{Martino, M.},
\bibinfo{author}{Claxton, R.} \& \bibinfo{author}{Strogatz, S. H.}
\newblock \bibinfo{title}{Redrawing the map of {Great Britain} from a network
  of human interactions}.
\newblock \emph{\bibinfo{journal}{PLoS ONE}} {\bf\bibinfo{volume}{5}}, \bibinfo{pages}{e14248} (\bibinfo{year}{2010}).

\bibitem{Ahn:2010uq}
\bibinfo{author}{Ahn, Y. Y.}, \bibinfo{author}{Bagrow, J. P.} \&
  \bibinfo{author}{Lehmann, S.} 
\newblock \bibinfo{title}{Link communities reveal multiscale complexity in
  networks}.
\newblock \emph{\bibinfo{journal}{Nature}} {\bf\bibinfo{volume}{466}}, \bibinfo{pages}{761--764} (\bibinfo{year}{2010}).


\bibitem{Bianconi:2009fk}
Bianconi, G., Pin, P. \& Marsili, M.
Assessing the relevance of node features for network structure.
{\it Proc. Natl. Acad. Sci. USA} {\bf 106}, 11433--11438 (2009).

\bibitem{Lloyd.:1982gf}
\bibinfo{author}{Lloyd, S. P.}
\newblock \bibinfo{title}{Least squares quantization in {PCM}}.
\newblock \emph{\bibinfo{journal}{IEEE T. Inform. Theory}}
 {\bf\bibinfo{volume}{28}}, \bibinfo{pages}{129--137} (\bibinfo{year}{1982}).
 
\bibitem{Scholkopf:1998lr}
Sch\"{o}lkopf, B., Smola, A. \& M\"{u}ller, K.
Nonlinear component analysis as a kernel eigenvalue problem.
\newblock {\em Neural Comput.} {\bf 10}, 1299--1319 (1998).

\bibitem{Tibshirani:2001lr}
Tibshirani, R., Walther, G. \& Hastie, T.
\newblock Estimating the number of clusters in a dataset via the gap statistic.
\newblock {\em J. Roy. Stat. Soc. B} {\bf 32}, 411--423 (2001).

\bibitem{park2007fast}
\bibinfo{author}{Park, H.}, \bibinfo{author}{Drake, B.}, \bibinfo{author}{Lee,
  S.} \& \bibinfo{author}{Park, C.}
\newblock \bibinfo{title}{{Fast linear discriminant analysis using QR
  decomposition and regularization}}.
\newblock \emph{\bibinfo{journal}{Georgia Institute of Technology, GA, Tech.
  Rep. GT-CSE-07-21}} (\bibinfo{year}{2007}).

\bibitem{Krebs}
\bibinfo{author}{Krebs, V.}
\newblock \bibinfo{note}{{http://www.orgnet.com/}}

\bibitem{Goh:2007fk}
\bibinfo{author}{Goh, K. I.}, \bibinfo{author}{Cusick, M. E.},
\bibinfo{author}{Valle, D.}, \bibinfo{author}{Childs, B.},
\bibinfo{author}{Vidal, M.} \& \bibinfo{author}{Barab\'asi, A. L.}
\newblock \bibinfo{title}{The human disease network}.
\newblock \emph{\bibinfo{journal}{Proc. Natl. Acad. Sci. USA}}
 {\bf\bibinfo{volume}{104}}, \bibinfo{pages}{8685--8690} (\bibinfo{year}{2007}).

\bibitem{Gursoy2000dgslg}
\bibinfo{author}{G{\"u}rsoy, A.} \& \bibinfo{author}{Atun, M.}
\newblock \bibinfo{title}{Neighbourhood preserving load balancing: A
  self-organizing approach},
\newblock in \emph{\bibinfo{booktitle}{Euro-Par 2000 Parallel Processing}},
  eds. \bibinfo{editor}{Bode, A.}, \bibinfo{editor}{Ludwig, T.},
  \bibinfo{editor}{Karl, W.} \& \bibinfo{editor}{Wism{\"u}ller, R.}, 
  Vol. \bibinfo{volume}{1900} of \emph{\bibinfo{series}{Lec. Notes Comput. Sc.}}, \bibinfo{pages}{234--241} 
   (\bibinfo{publisher}{Springer},
  \bibinfo{address}{Berlin/Heidelberg}, \bibinfo{year}{2000}).

\end{thebibliography}
\end{document}